\documentclass{elsart}
\begin{document}
\begin{frontmatter}
\title{Can magnetism-assisted quasiperiodic structures in Russell-FeS `bubbles' offer a quantum coherent origin of life?
}

\author{ Gargi Mitra--Delmotte\thanksref{gmail}}
and 
\author{ A.N. Mitra\thanksref{mmail}} 
\thanks[gmail]{gargijj@orange.fr}
\thanks[mmail]{ganmitra@nde.vsnl.net.in}
\address{244 Tagore Park, Delhi 110 009, India }

\begin{abstract}
 This paper seeks to expand the scope of  the  alkaline seepage site hydrothermal mound scenario of Russell et al, by appealing to its wider canvas for hypothesizing  self-assembly of gregite clusters  via interplay of forces within the gel phase of FeS membranes :  directed heat transport a la Rayleigh-Benard convection for dissociation,  vs oriented attachment ( small clusters) and  magnetic forces ( large clusters) for association, the latter assisted by magnetic mound constituents. The directed movement of tiny clusters through  cluster layers are reminiscent of  processes like budding, molecular motors, pre-RNA world on the lines of Cairns Smith's hypothesis, and optical polarity. Higher rate of (soft )  multinucleate formation vs growth rate of (rigid  )  microcrystals,  correlates with icosahedral (forbidden crystallographically ! )  framboidal  morphology. This pattern  indicates a link to phylotaxis, thus reinforcing the  quasi-periodicity connection which  can provide a natural access to features like surface limit, anomalous transport  and  low thermal conductivity, while facilitating diffusion through clusters.   And  magnetism offers  a  hierarchy of features :  primordial multicellularity;  phase correlations of  assembled molecules;  overcoming thermal decoherence.   These dynamical  nested structures  offer  possibilities for iterative computations, adaptive learning, and  coherent quantum searches. The link between enzymatic FeS clusters and the Hadean ocean floor is seen as part of a larger conceptual framework uncovering a role for Magnetism and the Origin of Life. \\

\end{abstract}
\end{frontmatter}

\section{  Introduction }

Wachtershauser drew a surface metabolism scheme where small molecular units could be indirectly surface-bound via association with  directly surface-bonded carrier constituents. This has the possibility of a `bucket brigade-like' drill between carrier consituents, that has a close parallel in transfer-reactions, partly preserved in extant biochemical pathways. Thus small units can be handed from one surface-bonded carrier to the next, even as they remain indirectly bound to the surface [Wachtershauser (1988)]. Moreover, the importance of ligand-binding and redox properties of transition metals in surface metabolism, a pivotal catalytic strategy used even today in enzymes, was emphasized by Huber and Wachtershauser (1997). These workers also demonstrated the formation, under simulated hydrothermal conditions,  of activated thioesters on co-precipitated FeS-NiS. This bears close resemblance to the (considered as ancient) acetyl-CoA pathway for CO$_{2}$ fixation where iron sulphide clusters constitute the catalytic core of the enzymes. But the main drawback of surface-metabolism based theories for the origin of life lies in their inability to provide a basis for concentrating the synthesized products. Once dissociated from the surface template, how could they be prevented from diffusing away into the vast expanses of the ocean? An ingenious mechanism resolving the concentration-compartmentation issues, was put forward by Russell et al (1994) who incorporated the `shopping requirements' of a number of origin- of - life theorists in their proposed iron sulphide compartments on the deep ocean floor of the Hadean Era, protected from the hostile primitive earth's surface environment. The interstices of the ocean-floor in more habitable, moderate settings, preferably not much above 40$^{\circ}$ C [Russell et al (1988, 2005)] but away from magmatic extrusions, provided the setting. Exothermic serpentinization of mafic and ultramafic (low silica content) oceanic crust by hydrothermal convection  of ocean water with contributions from magma degassing thus helped to richly load the escaping alkaline fluids with CH$_{4}$, H$_{2}$S, NH$_{3}$, HCN, HCHO  as they seeped out through the interstices to meet the mildly oxidizing, acidulous, CO$_{2}$ and Fe$^{2+}$ containing cooler waters on the ocean floor. Other than acting as catchment areas of nascently synthesized activated products, these precipitating FeS barriers comprising greigite and mackinawite, prevented the immediate titration of the two fluids and controlled their interactions. Four potentials ( hydrodynamic, thermal, chemical and electrochemical) existed across these membranes [Russell and Arndt (2005)], in tune with Nature's preference for being in the middle of extremes. Here, the mound, consisting of Mg-rich clays, ephemeral carbonates, green rust, as well as the sulfides, acted as a natural, self-restoring hydrothermal reactor. 
\par
These FeS membranes in the growing porous mound were seen as catalytic surfaces that also endowed the chambers with semi-permeability. Thus, newly synthesized compounds were captured, even as small uncharged compounds could diffuse through, with restricted electron and proton flow  through the mackinawite lattice [Russell et al (2003)]. The metal layers of the latter allowed electronic conduction,  while its sulfide rich layers acted as insulators against traversing the membranes. The membranes are thus seen as equipped with three main factors : a proton gradient, electron transfer agents, and energy-rich thioester bonds, that could  have driven primitive autotrophic metabolism [Russell et al. (1994)].  

\subsection{  Ancient roots of Iron-sulphide clusters}

The emergence of precursors to the present day clusters forming active enzyme centres [Huber and Wachtershauser (1997)] or `protoferredoxins' [ Russell et al. (2005)] can be envisaged naturally from such a hydrothermal mound. For even today this family of related iron-sulphur proteins continues to cater to a wide range of functions that, aside from taking advantage of their stabilizing influence in structural roles, essentially exploit the redox properties of these moieties as electron transfer agents, hydrogenation enzymes, redox catalysts, hydrogenases, dehydrogenases, nitrogenases, hydrolases, endonuclease III, in cytochromes, as redox sensors. A  notable example is the polyferredoxin from \textit{Methanothermus farvidus}, containing  ferredoxin-like repeats  that permits two- electron transfers across the membrane [Russell et al (2003) and references therein]. In this context, the presence of redox gradients in the 'Russell mounds' are seen as driving electron transfers across the protoferredoxin units. To boot, the 'mound scenario' appears to have greater chances of evolving preferences for optical isomers, say, as compared to an aqueous phase system, with it's proposed moderate temperature range, the FeS membrane catalytic surfaces and the aqueous chambers for entrapping and concentrating the released products. These are important ingredients towards the evolution of surface-synthesized products, with a tendency to use only one out of two optical isomers, a key feature of biosystems. 

\subsection{  The scale-free structure of the membranes}

The electron micrograph  of the iron-sulphide gel compartments  can be seen in Martin and Russell (2003) where the size of the chambers and walls vary around 20 microns and 5 microns, respectively. Diffusion controlled reactions would be expected to slow down with increasing thickness of aging sulfide membranes. As pointed out by Russell et al (1994) (while citing Kopelman's analysis [Kopelman (1989)]),  gels in general lie at the boundary of liquid and solid states, comprising a distribution of self-similar clusters over a range of sizes that are fractal on all scales. This facet is noteworthy, keeping in mind Nature's choice of a fractal geometry.
 
\subsection{ Interest in Greigite} 

On the Hadean ocean floor there would presumably have  been many alkaline seepage site hydrothermal mounds. Then, sensitivity to initial conditions could have generated variations in the composition of precipitating barriers. Energetically facile transformations 
could have lead to subtle composition changes with consequent ramifications on properties ;  two examples are i) transformation of   pentlandite, (Ni,Fe)$_{9}$S$_{8}$,   to violarite,   (Ni,Fe)$_{3}$S$_{4}$  under mild hydrothermal reaction conditions [Tenailleau et al (2006)], and ii) solid state transformations of griegite and mackinawite resulting  in ferromagnetic smythite overgrowths [Krupp (1994)].  
\par
Now the particular interest in griegite lies in its close resemblance to the clusters in the ancient pathway enzymes. This mineral forms as an intermediate in pyrite formation from mackinawite at high temperatures 150$^o$ C [Hunger and Benning (2007)]. Under the more mild temperature and alkaline conditions in the 'Russell mounds', the FeS membranes are envisaged as a composite of mackinawite (as a stabilizing frame) and griegite [Russell et al (2005)]. In fact, selected -area  electron  diffraction patterns of biomineralized crystals [Posfai et al, 1998] show the two minerals oriented relative to one another with $(011)_m //(222)_g $ and $[110]_m // [100]_g $.  This enables a parallel disposition of cubic closed-packed layers in both structures, embedded in a continuous sulphur-substructure maintained across the interfaces. 
\par
Greigite is the primary carrier of magnetization in many types of ancient sedimentary rock [reviewed in Posfai and Dunin-Borkowski (2006)]. The magnetic single- domain range for nanocrystalline griegite, forming from mackinawite, can be seen to vary from $\sim$ 50 nm to a wide upper range of  200--1000 nm ;   and the coercivity of crystals helps  contribute towards remanent magnetism of sediments [Diaz-Ricci and Kirschvink (1992)]. In fact  framboidal pyrite, ( named after its raspberry appearance ; framboise in french), is  the dominant morphological form  in ocean floor sediments.  Calculations by Wilkin and Barnes (1997) indicate that they owe their shape to magnetic accretion of the precursor greigite microcrystals whose morphology remains preserved upon conversion [Posfai et al (2001)]. Although framboidal pyrite has been shown to be synthesizable directly from FeS, the role of the former pathway seems to be relevant to the marine sediments where mixed greigite/pyrite framboids have been observed. Moreover, nucleation, growth and spheroidicity characteristics of pyrite framboids are thought to depend on shape inheritance in replaced spheroidal greigite,  as well as conditions within associated microbial mats [Popa et al. (2004)]. 

\subsection{ Icosahedral framboids }

Other than the spherical framboids, a highly ordered icosahedral type has been reported where this packing is maintained in its internal structure. The formational environment is evidently critical for the architecture of these structures that can vary from disordered to remarkably ordered arrangements of constituent crystals. Correlations between framboid diameter ($D$) and microcrystal diameter ($d$)  from modern sediments were investigated by Wilkin et al (1996). As pointed out by Ohfuji and Akai (2002), $D/d $ ratios of framboids dominated by irregular or loosely packed cubic-cuboidal microcrystals are low compared to high corresponding values observed for those composed of densely packed octahedral microcrystals. 
Interestingly, the latter variety were found in sedimentary rocks more than 11,000 years old [Roberts and Turner (1993)], where the central parts of the weakly magnetized framboids were found to have greigite microcrystals. Sections from these show that the pentagonal arrangement comprise a central pentagonal domain with its sides connected to five rectangular/trapezoid-like regions  which are in turn connected via fan-shaped domains. The arrangement pattern of these densely packed octahedral microcrystals linked edge to edge is `lattice-like' (space filled) in the rectangular domains whereas in the triangular domains  the triangles are formed by the ($111$) faces of the octahedral microcrystals and the voids between them. Thus within these domains the individual faces of the microcrystals do not make any contact. The icosahedral form is seen as generated by stacking twenty tetrahedral sectors  packed on three faces out of four, and connected by their apexes at the centre. Generally acknowledged as dynamically stable, this form is known to have six 5-fold axes at each apex,  and ten 3-fold axes at each face, as can be seen in a number of naturally occuring structures  from microclusters like fullerene to some viruses [Ohfuji and Akai (2002)].

\subsection{ Outline of paper }

\begin{enumerate} 
\item We hypothesize the self-assembly of greigite clusters within the FeS membranes envisioned by Russell et al, via an interplay of forces based on evidences from condensed matter mesoscopic energy dissipating systems and the present understanding of the gel-state. Directed diffusion due to convection currents are seen vis-a-vis biological phenomena like budding, molecular motors, pre-RNA world (a la Cairns-Smith's hypothesis), and optical polarity.
\item We posit the nature of these greigite aggregates as scale-free quasiperiodic ordered structures in the `Russell membranes', based on well-known observations.
\item We look at these hypothetical quasi - periodic structures  in the light of their potential biological implications;  also their potential as ancestors to biosystems with many facets such as directional order, and feedback-based spatially asymmetric organization. 
\item We review possibilities arising out of quasiperiodic order and ask if these magnetically ordered clusters could provide a coherent connection between quantum and macroscopic classical realms. These structures are considered as possible ancestors of life, with a capacity for quantum information processing. The myriad manifestations of magnetism, especially its significance in biology , are briefly reviewed in the light of these clusters towards invoking a possible magnetic origin of life. 
\end{enumerate}

\section{Self-assembly of greigite and diffusion through clusters} 

\subsection{ Magnetism of mound constituents}

In this paper we would like to draw attention to a key feature of one or more constituents comprising the base of the hydrothermal mound : that of \textit{magnetism}. Particularly ferrimagnetic species like griegite retain magnetic remanence upto temperatures of 300--350$^{\circ}$ C [Hu et al (2002)]. The formation of greigite (Fe$_{5}$NiS$_{8}$) in the iron-sulphide `bubbles' proposed by Russell et al, would be taking place in a mound where magnetic field lines would be concentrated due to the magnetic components forming the base of the mound, a process dating back to primary crustal formation when the molten `compass-like'constituents cooled from magma, freezing in the geo-magnetic field direction. According to a recent SQUID (superconducting quantum interference device)
estimation of laser heated ancient crystals of feldspar and quartz, the Earth's magnetic field was at least half as strong 3.2 billion years ago as it is today, thus hinting at the possibility of the rotating and convecting iron inner core having formed well before 3 billion years ago [Tarduno et al (2007)]. 

\subsection{ Self-assembly in ferro-fluid mesoscopic systems} 

We hypothesize that the magnetism of the mound constituents is likely to have assisted the formation of greigite from mackinawite in the freshly precipitating FeS membranes in the `Russell mounds' and the consequent alignment of the nanoparticles into self-assembled structures. Experiments closest to proof of concept, are studies on ferrofluids containing colloidal suspensions of thermally disoriented single domain nano-particles, with their individual moments proportional to their volumes. In these a magnetic field assisted assembly under energy dissipating conditions, has been recently demonstrated [Ganguly \& Puri (2006)] as a result of the interplay of magnetic forces with others : surface tension, fluid shear or gravity. In particular, the competition between combined advective-diffusive hydrodynamic forces with magnetic forces in these experiments with ferrofluids, are extremely pertinent to the phenomena envisaged by Russell et al in the hydrothermal mounds. The magneto-crystalline anisotropic energy component that underlies rotation of a magnetic particle with finite moment, and aligns it to the external magnetic field by competing against its thermal energy component, is a function of the particle size, shape and structure. And, microscale heat transfer via thermomagnetic convection phenomena are exactly what one can expect of nano-sized magnetic particles as gravitational effects are not significant on these length scales. Along these lines [Finlayson (1970)], the temperature gradient across the mound can be expected to generate a gradient in the ferromagnetic susceptibility of the freshly forming FeS colloids, and in the presence of the magnetic field due to the mound base, this varying susceptibility would lead to a modulation of the effective magnetic force field with consequent thermomagnetic convection of the nanoparticles. In fact, the recently demonstrated possibility of microstructure control in a phase-selective assembly of ferrosulfide microrods, via an applied magnetic-field in a hydrothermal process [He et al (2006)] underscores the magnetic-tunable properties of ferrosulfide building blocks. 
\par
We must bear in mind that these experiments differ from the `Russell context' in that a gel medium characterizes the latter unlike liquid carrier dispersed magnetic colloidal particles used for e.g. in [Ganguly and Puri (2006)], that are stabilized via coating etc., against aggregation. The gel-like membranes would also  provide an environment allowing motional freedom intermediate between regular solid and liquid states. 

\subsection{ Interplay of forces within membranes}

In contrast to investigated ferrofluid systems, the growing porous mound is seen as comprising membranous FeS compartments with entrapped fluid. The difference in ferromagnetic susceptibility of particles across the gradient spanning the mound (order of cm at the fresh growing end of structures of the order of meters) would obviously be several decades higher than that across the gel-phase of individual bubble membranes (microns). At the growing front,  the low temperature end of the gradient that is, the susceptibility would presumably be high. Now the thermal gradient would be expected to transfer heat via fluid motion a la Rayleigh-Benard convection. Thus heat would be transported through the entrapped aqueous spaces as well as  the  bubble membrane layers : A network of scale-free hydrated clusters  characterizing the gel phase (see above)through which thermal energy  would be pumped  in.  This can be expected to break bonds, permit motional freedom (e.g. rotation of freed clusters) upon which re-associations would presumably result in better alignment among clusters. Elimination of interfacial water and crystallographically coherent alignment of coalescing clusters are seen as steps towards nucleation and growth processes [Navrotsky (2004)]. Note that structure and bonding in clusters have several features in common with their  crystalline mineral counterparts. Thus, in this manner oriented attachment may have helped build up cluster size. The transformation from mackinawite to greigite could have occured at this stage or earlier. In any case greigite clusters below the size of of 30-50nm are thought to be superparamagnetic, with single domain oriented arrays forming above that size (see subsection 1.3), thus pointing to the importance of magnetic forces  [see Wilkin and Barnes (1997) and references therein]. 
\subsection{ Inspiration from Cairns-Smith's clay hypothesis }
On the lines of  'crystal propagation' envisaged by Cairns--Smith (1985), we draw a parallel with self-assembled greigite clusters, helped by the magnetic constituents comprising the mound base, in overcoming thermal forces as outlined above. Such an ordered arrangement of greigite molecules embedded in mackinawite lattices [Russell and Hall (2006)]  could well have been a start-point for offering nanoscale aperiodic and catalytic surfaces within these structures. Significantly, the importance of an aperiodic structure for holding information was pointed out by none other than Erwin Schroedinger (1944) several decades ago. 
\subsection{ Self-replication of nested clusters; self-production of greigite }
In contrast to the growth of conventional crystalline lattices restricted to a growing surface, the directed diffusion of tiny clusters (in proportion to size) would occur through intra-layers of large clusters. Field-assisted alignment of greigite units (within the gel-phase of the 'Russell bubbles') would not only be expected to lead to alignment on the cluster surfaces, but diffusion (see Table) would also allow these units access to topologically equivalent templates offered by smaller scale nested structures. Each of these in turn would therefore have the potential of inflating into larger nested clusters themselves, analogous to `self-pattern' conserving and replicating, budding structures, ubiquitous in biological systems. 
We also note that this scenario closely resembles the bigger scale phenomenon corresponding to the parent FeS membranes, envisaged as growing FeS 'bubbles' [Russell et al (1994), Russell and Hall (2006)].
\subsection{ Primitive molecular motors}
Similarly, there is a striking topological analogy to the gradient-assisted diffusion of loaded greigite units through intra-cluster layers with the directed movement on aperiodic intracellular surfaces displayed by bio-molecular motors. Looking at the machinery crucial for cellular functioning (the translational, transcriptional, cytoskeletal assemblies) one can identify an invariant and simple topological theme: The movement of a cargo loaded element on a template or scaffold (representing a varying potential) that harvests thermal fluctuations for dissociating its bound state and spends energy (usually ATP currency) for conformationally controlled directed binding, or an ionic gradient for guiding the direction [ Astumian (1997)]. We make an analogy of this ratchet system with a greigite element liganded to a newly synthesized molecule in the `Russell chambers', diffusing through a cluster of dynamically assembled greigite elements assisted by magnetic constituents in the 'Russell mound'. This `cluster propagation' on the lines of Cairns- Smith's `crystal propagation' involves an interplay of thermal fluctuations with the oriented assembly of aligning magnetic elements (binding via weak forces) that get propelled in the direction of the thermal gradient. Thus we foresee that the diffusion of the loaded greigite molecules would have been propelled in the direction of the thermal gradient described by Russell and Arndt (2005) due to the aperiodic potential of the greigite mineral composing the ordered clusters, in the process impregnating these 'micro-compartments' with the seeds of primitive biomolecular motors. All this emerges from the fact that random Brownian forces are the right magnitude (average thermal energy $1k_BT$ approx 4.1 pN nm at NTP) for driving nano-sized particles [see Phillip and Quake (2006)], while inertial effects are negligible. But note that 'biased' Brownian motion, based on structural anisotropy alone, is not permitted by the Second Law of Thermodynamics; so to overcome this problem, Feynman had proposed the application of a thermal gradient !
\subsection{ Pre-RNA world}
We now turn to the proposed RNA world [see a recent review by Orgel (2004)], where RNA played the roles of \textit{both} DNA and protein --- for convenience call them RNA---sequential and RNA-structural, respectively. Evidently, nature designed DNA for packaging information efficiently, satisfying Claude Shannon's maximum entropy requirement, as evident in the lack of correlations across the sequences. This is precisely what compounds the resolution of the 'chicken-egg' conundrum, as the largely random nature of sequential information encoded in the DNA is correlated via RNA with the high degree of stereo-chemically defined structural information in proteins. (This is not to undermine the role of structural signatures in DNA, say for its sequence-structural recognition in regulating transcription, but as compared to the structural information content of proteins, one can say it is relatively more limited). Now, the paradox of the appearance of RNA-sequential may be easier to resolve if we envisage that it's appearance in the first place was due the replacement of a structurally less specific interaction. By analogy with Wachtershauser's `bucket brigade-like' transfer reactions (here his `carrier surface' (see section 1) is the `underlying manifold' (see  5.1.3) due to the interplay of `dissociative' and `associative' forces), it is plausible that the intra-layers of large greigite clusters offered themselves as 'templates' for directed diffusion (see above) of tiny greigite clusters loaded with activated products synthesized in the 'Russell compartments'. These `templates' could themselves be envisaged as a primitive translational machinery, with oriented greigite units playing the key adaptor roles \textit{a la} transfer RNAs. Here one surface (carrier) is part of a magnetically oriented array, with limited structural information, while the other face is liganded to activated compounds rich in structural information that were synthesized in the mound itself, or brought in from elsewhere by diffusion. Admittedly, this `magnetic letters-like' scenario bears a striking resemblance to the tRNA's bringing the amino acids together for stringing them up on the basis of the sequential information inscribed in the mRNA template. With the passage of time, as visualized in the RNA world scenario, 'RNA-sequential' might have replaced the magnetic template, while exploring catalytic properties in the released/unbound 'RNA-protein' form.
\subsection{ Possible origins of optical polarity}
We saw above that the presence of a thermal gradient would have caused directed diffusion of activated synthesized molecules riding 'piggy-back' (ligand-bound) on the diffusing and aligning greigite elements; these products would be linking up on the templates provided by the intra-layers of the dynamically assembled clusters. This directional asymmetry may well have pushed the balance in favour of bond formation between juxtaposed activated units having the same chirality close to the ligand-binding site, perhaps even aided by the space constraints of such intra-layer activity. Thus, the optical activity of the first-bound unit -- the symmetry-breaking choice -- might have set the preferences for those of the consequently selected units from the soup of chemicals waiting in the aqueous interiors of the 'Russell compartments'. In this manner, over time, the directed diffusion on template layers of the clusters may have been a likely scenario for evolving preferences in favour of one optical isomer over its counterpart, observed in biosystems today. 
\subsection{How about a surface limit ?}
The above scenario  is envisioned to occur within the gel phase Russell membranes which themselves demarcate an upper limit for the protocell-like clusters.  As discussed later ( see 4.5 - 4.7  ) the gel phase  is endowed with  features  resembling biological order ( a connected network, nested hierarchial organization ).  All this   begs an insight as to whether a more precise resolution of the surface limit can be envisaged for the assembled clusters.  .  Towards that end, we look at observed structures  of icosahedral framboidal greigite. 
\section{ Hypothetical quasi-periodic structures} 
After browsing quickly through a few facts from the literature, which \textit{apriori} may not appear connected, we shall attempt to propose a hypothetical structure for these clusters, possibly consisting of griegite.
\subsection{ Icosahedral framboidal greigite}
In an investigation of apparent biologically induced mineralization by symbiotically associating bacterial and archaeal species, framboidal greigites have been obtained from Black Sea sediments that are ordered clusters of octahedral crystals comprising Fe$_{3}$S$_{4}$-spinels (Essentially cubic where sulphur forms a fcc lattice with 32 atoms in the unit cell, and Fe occupies 1/8 of the tetrahedral and 1/2 of the octahedral sites). Their size is restrained by their icosahedral symmetry and under greater pressures at depths of 200m, the diameters are mostly $\sim$2.1, $\sim$4.2, 6.3 or 8.4 microns, with the two intermediate ones predominating. The smallest of these are formed from 20 octahedral crystals (0.35 micron) positioned at the apexes of an icosahedron and surrounding a 0.5 micron diameter vacancy that give rise to 12 pentagonal depressions on the outside. Nested structures building up from this smallest one lead to the higher sized clusters [Preisinger and Aslanian (2004)]. 
\subsection{Quasicrystalline order---a brief introduction}
Quasicrystals [Trebin (Ed.) (2003); Suck, Schreiber, Häussler, (Eds.) (2004) ; also see quasicrystal tutorial websites maintained by Uwe Grimm, Steffen Weber] are aperiodic structures with perfect long-range order but also having classically forbidden rotational symmetry elements, incompatible with translational periodicity in 3D space. They are like crystals in giving sharp, discrete diffraction spots, but at the same time also like amorphous solids in which no unit cell can describe the atomic packing. Therefore, in contrast to conventional crystals, where three Miller indices can be used to label the diffraction intensities of translationally periodic structure, at least 5 independent vectors are required for quasicrystals: 5 indices for polygonal and 6 for icosahedral quasicrystals. To get these generalized Miller indices, one starts with the requisite n vectors forming a nD-reciprocal space. The corresponding nD-direct space generating the diffraction pattern can be regarded as the hyper space that provides a periodic description of the quasiperiodic structure and which can be suitably projected to get the structure in 3-D space. Again, in contrast to classical crystals whose symmetry group can be associated with one of the 230 space groups for procuring the structural description, there is a lack of a corresponding non-trivial group of isometric transformations that can leave quasicrystals invariant. Nevertheless, this is thought to be more than compensated by the presence of self-similarities so that the structure remains invariant under affine homothetic (scaling) transformations.
\par
Multi-fractal or critical eigenstates lie in between the spatially extended eigenstates in conventional lattices which lead to ballistic movement of electrons like in metals, and exponentially localized eigenstates in strongly disordered systems that result in insulating phase with zero mobility (Zhong et al 2000). Clearly, the insulating behaviour of composite metallic elements is a contrast to that of conventional alloys. This atypical transport is generally linked to the presence of a broad pseudo-gap, that is revealed from the reduced DOS (electronic density of states) profile at the Fermi level (EF). Presumably, the presence of a matching distance occurring repeatedly in the structure leads to the energy lowering, resonant scattering of the conducting electrons at EF [see Haberkern (1999)]. Apart from critical eigenstates and multifractal DOS profiles, hopping mechanisms between clusters have also been proposed among schemes to explain such transport behaviour. These are seen as induced by interband transitions between electronic eigenstates of bonding electrons, showing spatial power law dependences. To that end, inflation rules take into account the self-similar assembly of atomic clusters in perfect icosahedral quasicrystals for obtaining the energy levels of invoked electronic supershells in clusters. These permit only certain values for composition and atomic valences with the fractal geometry defining the sites for recurrent localization of bonding electrons that are distributed into `magic' cluster states, towards stability of the shell structured framework [(Janot (1997)]. In another study, the anomalous power law temperature dependence of conductivity was used to differentiate between the two low-temperature conducting regimes, observed for quasicrystals (Roche and Fujiwara (1998)], applying the theme of quasiperiodic mesoscopic order as the basis for generating coherence. 
\par
In icosahedral materials, atomic clusters seen as recurring structural motifs, usually comprise nested polyhedra of several tens of metal atoms. Localized properties of atomic clusters (see above) do contribute towards some exotic physical properties. Thus, associations of properties that were never observed, for example in conventional metallic alloys, can be seen in densely packed quasicrystalline materials: very hard, fragile and also extremely poor conductors of heat and electricity; they are resistant against oxidation while also showing anomalous surface properties such as reduced wetting (high contact angles with water) and low friction coefficients [reviewed in Thiel (2004)]. In fact, resistivity is seen to increase concomittantly with improved sample quality, contrary to classical crystals where quality levels correlate with lesser scattering centres and reduced resistance. These and other unusual features, such as diffusion while maintaining long range order, optical properties, have turned them into promising candidates for a range of technological applications: as coatings, metal matrix components, hydrogen storage materials, thermal barriers, infrared sensors, alternatives of photonic crystals, etc.
\par
Sure enough, the interwoven patterns characteristic of quasicrystalline organization of nested self-similar structures, cannot be found in conventional lattices with long range translational symmetry of allowed symmetry elements [Coddens (1991); Kostov and Kostov (2006)]. Indeed, their physical properties in unusual combinations seem to have important implications for biological processes. Some of these are enlisted in the Table. It must be kept in mind that these descriptions are based mainly on (faceted) solid phases of metal alloys, in contrast to the present context of assembling molecules  within the gel-medium of the 'Russell membranes'. 
\begin{table}
\caption{Potential implications of quasicrystalline order in bio-processes}
\begin{minipage}{\textwidth} \centering 
\begin{tabular}{p{6cm}p{7cm}} \hline
{\em Quasicrystalline order property\footnote{ see Roche and Fujiwara (1998); Suck, Schreiber and H\"aussler (Eds.) (2004); Thiel (2004); Trebin (Ed.) (2003)} }& {\em Implication for bio-processes } \\ \hline
very low friction coefficients & useful for diffusion, keeping long range order
and 
storage \\
hard and brittle and contain catalytic amounts of active metals & cleavage 
and catalysis \\
resistance against oxidation and reduced wetting & protection 
and useful hydrophobic surface property \\
very low thermal conductivity & maintain low temperature and assist coherent processes \\
atypical electron transport & non-ballistic quantum diffusion (long relaxation times) 
hopping mechanisms (shorter times) \\ \hline 
\end{tabular}
\end{minipage} 
\label{table}
\end{table}
\subsection{ Transition metals} 
The effects of transition elements with regard to the electronic and transport properties of quasicrystals are well known, particularly the role of strong sp (from Aluminium) to d coupling (due to an ordered sublattice of transition metal atoms) in pseudo-gap formation. The temperature dependencies of the experimental resistivity of quasicrystals, were compared to that of amorphous and metallic crystals in a temperature range of 4 to 300K. For the stable quasicrystals (i-AlCuFe and i-AlPdMn), the range of resistivity was $\sim 10^{4}$ near $T = 4$, decreasing to $\sim 10^{3}$ at $T = 300$. This was in contrast to metallic alloys where the resistivity increased from less than one at $T= 4$K to less than ten at $T=300$K [Trambly de Laissardiere and Mayou (2007)]. 
Also, as mentioned above, the d-orbitals of transition metal ions, play a key role in surface metabolism based theories for the origin of life. According to Huber and Wachtershauser (1997), all chemical conversions of primordial metabolism occurred in a ligand sphere, held together by bonding to the surfaces of iron-sulphur minerals, where transition metal ions e.g. Ni$^{2+}$, Co$^{2+}$, are catalytically active. 
\subsection {Hypothesis of quasi-periodically ordered clusters}
We draw inspiration from the following: 
\begin{enumerate}
\item Quantum coherent search processes in biology [see Davies (2004a)] and demonstration of coherent alignment of magnetic dipoles induced over an actin filament contacting myosin elements in the presence of ATP [Hatori et al (2001)] that indicated the potential role of magnetism, such as in overcoming thermal decoherence effects, keeping in view the remanence of sedimentary rocks. 
\item Observations of fractally ordered structures frequently in Nature [ Selvam (2004)] and experiments on quasicrystalline, icosahedral phases [e.g. Abe1 et al (2004); Shinjo et al (1986); Fan et al (1990); Belin-Ferré (2004)], in dynamical self-assembly [e.g. Domrachev et al (2004)] and the potential implications of such quasiperiodic order in nanoscale biological processes as given in the table

\item Age-old speculations of framboids as biological fossils and discovery of icosahedral framboids of greigite in Nature.
\item Russell's suggested description, quoting Kopelman's (1989) analysis of gels ( see subsection 1.2 ), of the membrane as having self-similar structures of different sizes, fractal on all scales [Russell et al (1994)], and the gel consistency of the cell. Also the cytoskeletal framework, the layered membrane organization of the cell, are intriguingly analogous to a nested quasiperiodic organization.
\end{enumerate}
The first set of observations seems to resonate with the idea of magnetically ordered clusters, describing the ancestor of Life. The second set of observations concerning nested quasiperiodic order has a link to the generic phenomena of Phyllotaxis (see 5.1.2), which in turn seems to be very relevant for point 3), i.e. the discovered icosahedral forms of framboidal greigite. The patterns of parastichies observed ubiquitously in nature, e.g. pinecones, pineapples, etc.  seems to strongly match with framboid patterns. As pointed out [Ohfuji and Akai (2002)] from the high D/d values of icosahedral framboids with densely packed octahedral microcrystals (see 1.4 ), this morphology could be related to the initial nucleation rate and growth time of microcrystals, and the number of microcrystals per framboid. In fact, the characteristic framboidal architecture with same size and shape of constituent microcrystals,  instead of a whole block of crystalline material, is in  itself  intriguing since the icosahedral form is known to be crystallographically forbidden. Further, while the process of crystallization seems to be in line with aging/hardening, giving way to death-like static structures, the contrasting formation of multiple nucleates  appears to be more in tune with life-like, soft, dynamic processes. The fourth point emphasizes the feature of self-similarity and the importance of the gel-phase. Note the recent interest in soft quasicrystals [Lifshitz \& Diamant (2006)]. 
\par
 And we therefore posit the presence of quasiperiodic, hierarchially arranged self-similar structures, composed of magnetically ordered molecules, such as of greigite, and forming by self-assembly within the soft `Russell iron sulphide membranes', as the origins of primitive cells that had molecular machinery. Being embedded structures, they are even more protected, while at the same time enjoying the benefits of the gradients and locally enriched (via entrapment of synthesized `metabolites' in the `Russell FeS chambers') shielded environment of the hydrothermal mound, proposed by Russell [Martin and Russell (2003)]. 
\par
As to the morphology of the hypothesized quasiperiodic structures embedded in the `Russell membranes', being envisaged as primitive cells, one possibility could be an icosahedral-like appearance, essentially inspired by the observations of the apparently stable geometrical forms of biomineralized framboidal greigite [Preisinger and Aslanian (2004)] and nested microcrystal arrangement in discovered fossils [Ohfuji and Akai (2002)], mentioned above. Further,  quasicrystalline phases are known to  co-exist with closely related micro- crystalline phases (see 4.3 below). Thus, the presence of greigite microcrystals and their nested organization, needs to be studied in the light that they may  possibly have originated from a quasiperiodic structure. 
\par
Aging would be expected to work on the composition and arrangement in the nested quasiperiodic ordered array of greigite, giving rise to stable crystals (similar to those observed). In this manner, they bow out of the growing dynamical phase and get engulfed in rigidifying death-like static, stable structures. In fact, as pointed out by Russell et al (1994), the leading surface of the membranous sulphide mound was always young as the freshly precipitating bubbles had the capacity to grow. According to their hypothesis, organosulphur ligated clusters in probotryoids along with other absorbed hydrophobic organics would be expected to at least inhibit if not block the crystallization of unligated clusters beyond greigite. However, as they `aged' these changed into more rigid and thickened walls which were no longer capable of expansion. 

\section{ A Quasi-Periodic Ancestor ?}

Assuming the hypothesized quasiperiodic structures are indeed formed in the 'Russell iron sulphide membranes', let's consider their potential as protocells.
\subsection{Diffusion through intra-cluster layers}
In Section 2, we saw that scenarios resembling budding, molecular motors, pre-RNA world and optical polarity could emerge from the directed diffusion of tiny greigite clusters through intra-layers of large assemblies. Now, quasicrystalline order has been found to be associated with a low friction coefficient (see Table). This feature should   facilitate diffusional transport through the intra-cluster quasiperiodic ordered layers. Note that in the context of budding like phenomena, a nested architecture also provides a natural basis for heralding in a hierarchial order - a key feature of living systems. 
\subsection{Surface-limit, diffusion, catalytic features} 
Quasiperiodic stacking of infinite non-identical atomic layers typify the bulk structure of quasicrystalline materials, making it hard to ascertain the surface layers. This is in contrast to the bulk planes with maximum atomic density defining stable conventional crystal surfaces, a rule that cannot be applied to the former which are packed on a different basis. An inter-layer spacing rule has been recently proposed [Sharma et al (2004)], based on scanning tunneling microscopy of the pentagonal icosahedral surface of the Al-Cu-Fe quasicrystal and the refined structure model of the iso-structural i-Al-Pd-Mn, towards explaining the stability of quasicrystal surfaces. These are henceforth seen to form as a consequence of bulk truncations at positions where blocks of atomic layers are separated by larger interlayer spacings. If these properties could be extended to the hypothesized griegite clusters, they would be endowed with similar surface-limiting features, in addition to protective surface properties: being resistant against wetting, oxidation, etc. (see Table).
\par
Furthermore, recent interest in quasicrystalline substances as potential catalysts emerges from the collective properties of a brittle nature and the fact that a number of Al-based quasicrystals contain minor amounts of catalytically active metals like Ni, Pd, Fe, and Cu [Sharma et al (2005)]. Thus, taken together with the low inter-layer frictional forces allowing diffusion and storage of diffusing molecules [reviewed in Nakajima and Zumkley (2001)], the likely association of quasiperiodic griegite structures with such catalytic features seem to be tailor-made in the context of biological phenomena.
\subsection{ Probable electronic conduction profile}
Barring crystalline approximants and quasicrystalline phases co-existing with twins via orientational correlations, the presence has been reported of quasicrystalline phases that can form by phase transitions of microcrystals existing with special orientational and phase relations as coherent domains. These are therefore capable of producing an icosahedral diffraction pattern and offer good starting points for quasicrystalline structure determinations [Coddens (1991)]. In a similar vein, as described above (see subsection 1.4) for the greigite-microcrystal containing core of the ancient sedimentary rocks, the lattice-like space filling arrangement exists solely in the five rectangular/trapezoidal domains of the densely packed octahedral microcrystals. Further, these are connected to each other only via triangular/fan-shaped domains within which individual faces of the microcrystals do not contact each other. The predominance of edge to edge contacts would possibly lead to poor electronic as well as thermal conductivity in such an architecture. The only possibility of classical conduction of electrons, if any, can be visualized along the five-fold symmetric planes where the octahedral microcrystals in the rectangular domains contact their triangular counterparts. From this picture, it appears that the closely related hypothetical clusters may have had anomalous transport properties like those of quasicrystalline ordered materials (see Table). Thus, electron transfers due to redox gradients across these non-conducting structures could have been enabled by a series of short tunneling currents, intriguingly somewhat like those seen in evolutionarily conserved iron sulphur electron transfer proteins [Daizadeh et al (2002)].
\subsection{Cleavage profiles}
In contrast to conventional periodic crystalline lattices, quasicrystalline structures are highly fragile (see Table). Thus, the potential cleavage capacity of these growing structures looks almost tailor-made for evolving and giving rise to biological phenomena. Indeed, the cleavage/breakage of mesoscopic, gel-phase, quasiperiodic structures growing by self-assembly, not only resembles processes like cell division, but also opening \& closing mechanisms observed for densely packed barrel shaped `vault particles'. The latter are suspected for storage (see Table) and intracellular transport in species, ranging from slime-molds to humans [Herrmann et al (1997)]. Besides, the redundant, distributed organization of this self-organizing system allows for a capacity to replace or `heal' cleaved/fractured domains. Presumably, this would also involve the feature of `self-production' as the formation of greigite units from mackinawite would be expected to proceed in parallel, being assisted by the magnetic field. 
\par
Now, biological systems display a variety of modes of copying information. If nucleic acids follow a template basis for assembly, membranes grow by extension of existing ones, ensuring reproduction of composition. The duplication of entire structures (e.g. spindle pole body or the dividing cell as a whole) leads to daughter structures conserving the parent structure patterns. Apparently, the golgi apparatus duplicates by matrix or template based self-construction, in contrast to the one in \textit{Toxoplasma gondii} which splits transversely into two [Harold (2005)]. Looking at all these possibilities, a prebiotic ancestor with a quasiperiodic lattice generated by dynamic self-assembly, seems to have the potential for explaining the subsequent evolution of such diverse modes. 
\subsection{ Possible origins of connected spatial organization }
The spatial organization of the cell or positioning of its elements remains an elusive feature that begs an understanding of how it all came about. As is increasingly becoming evident, genetic information constitutes only a fraction of the inheritance. For, the progeny are modelled on the 'parent template' that provides the precise spatial information for element organization and patterns, at different levels of the intricately connected hierarchy [Harold (2005)]. Irrespective of species, the cytoplasm appears to have rich structure, with increasingly reported associations of mobile proteins with defined, albeit transient, locations. Biomolecules like proteins and ions play a critical role in structuring of intracellular water [Chaplin (2006); see Martin Chaplin's website]. Like hydrated cross-linked polymer gels, the cytoplasm thus exhibits excluded volume effects and sizeable electrical potentials. Actually, the cell is viewed as a gel; gel-sol phase transitions underlie its dichotomy that can be accessed via subtle environmental variations leading to finite structural changes [Pollack (2001)]. 
\par
The latter, taken together with the connected, fractal cellular organization, prompts us to look for its possible origins in the scale-free, structured gel-phase of the Russell iron-sulphide membranes. This is because firstly, like the cell, the FeS membranes provide an order intermediate between aqueous and solid phases. Secondly, there is immense interest in the fractal properties of quasiperiodic order that lies in the midst of disorder in amorphous systems and the perfect periodic order of crystals. And the cell abounds in scale-free structures: from the nested cell structure to the highly networked cytoskeleton, an order that persists at progressively smaller scales. But here, although much work has been carried out on quasiperiodic clusters in alloys, the corresponding information on ordering in gel states is yet to come. Nevertheless, a \textit{magnetic feedback} basis for ordering would underlie the long range quasiperiodic organization in these dynamically generated clusters. Undoubtedly, the paradigm of self and non-self interactions between cellular components leading to coherent cooperative dynamics [Ling (2001)], seems to be well in line with the possibility of such a coherently connected ancestor nesting within the 'Russell membranes'. 
\subsection{ Magnetic ordering in hypothetical 'proto-cell' clusters}
Next we look at the nature of magnetic order in some icosahedral clusters [Estiu and Zerner (1994); Kashimoto et al (2006)]. Here theoretical predictions have had mixed responses from experiments. For although at a first glance, aperiodicity might be seen as provoking geometrical frustration, a number of theoretical models concerning magnetic ordering in quasicrystals indicate long range order. The discovery of icosahedral Zn-Mg-RE quasicrystals (RE represents one of the rare earth metals Y, Gd, Tb, Dy, Ho and Er) was followed by contradictory claims for and against long-range antiferromagnetic order in these quasicrystals where only the presence of short-range spin correlations appears non-controversial [Lifshitz and Even-Dar Mandel (2004) and references therein]. The situation is clearer for yet another group of newly discovered Cd-Mg-Tb icosahedral quasicrystals with comparitively less local frustration facilitating coupling between neighbouring spins. In addition, negative Weiss temperatures indicate dominant antiferromagnetic interactions between the RE magnetic moments [Sato et al (2001)]. Also, symmetry considerations point towards an antiferromagnetic ordering in a perfect icosahedral structure [Vekilov et al (2005); Lifshitz (1998)].
\par
How does all this relate to the hypothesized clusters? Dynamical constructions with a constant flux of elements [Harold (2005)] in and out of the clusters (in a gel like environment) are unlikely to be hampered by local frustrations of magnetic ordering that are possible in static structures. (This is not to suggest that geometrical frustrations will not occur in such clusters [Vedmedenko et al (2006)] ). Logically, the antiferromagnetic ordering of a newly formed perfect icosahedral structure should even aid in its 'budding out' of the parent cluster. However, the maintainance of perfect symmetry of these structures in a gel-like membrane environment seems unlikely with dynamical accompanying phenomena, such as budding processes, presumably inviting distortions. More importantly, 'piggyback' diffusion phenomena on aligning magnetic molecules, etc. would obviate the presence of other molecules leading to screening effects, in proportion to the complexity of the evolving structures. Hence it is quite likely that the growing individual clusters were initially associated with a finite magnetic dipole moment; this could have provided a primitive basis for their linear (pole to pole) alignment (see 4.8).
\subsection{ Nested arrangement of spatially oriented structures }
Biological structures appear as nested organizations based on coherent feedback [Ling (2001)] through a lattice of interacting, spatially oriented units. Even intracellular water is structured. In fact, 'site-dipoles' have been proposed for resolving the apparent contradiction between the seemingly random molecular movements and the correlated orientations in assemblies. Thus the co-operativity among water molecules occupying the site-dipole field surrounding a solute in MD simulations, manifested in coherent patterns ($\sim$14A$^o$) that lasted about 300ps, even as individual molecules randomly moving in and out of the sites, rapidly lost their orientational memory [Higo et al (2001)]. For an insight into the possible nature of structuring, Chaplin's model (2004) considers the K-carboxylate ion pairs that invite surrounding clathrate water structuring. This in turn leads to icosahedral water structures that ensure maximal hydrogen-bond formation while cluster mobility helps in searching out cooperative H-bonding partners. (An important fallout of such feedback-based reinforcement of tetrahedrally ordered structuring, with pentagonal symmetry, is its increased resistance to freezing). On higher scales, the directionality of biochemical processes gets derived from the asymmetric structure of biomolecules and their association into consequently polarized assemblies with increasing complexity [Harold (2005)]. These observations have a striking analogy with spatially oriented magnetic elements settling into similar nested structures. Their magnetic moments endow them with spatial orientation, while a quasiperiodic arrangement allows nested long range order in clusters. 
\par
We also note that some fundamental biological structures form from asymmetric monomers. For instance, the directionality of nucleic acid polymers stems from the asymmetry of template-based aligning monomers. The cytoskeletal family of proteins provide another outstanding example. The past two decades revealed how analogous functions are carried out by bacterial homologues of eukaryotic cytoskeletal proteins. Actually, the highly conserved FtsZ, barring a few exceptions, is found across all eubacteria and archaea. Despite it's low sequence identity to tubulin, it's eukaryotic homologue, the two proteins not only share the same fold but follow similar self-assembly patterns, forming protofilaments. The longitudinal contact of the assembling monomers is in a head-to-tail fashion. The other crucial eukaryotic cytoskeletal protein - actin - also shows a distinct asymmetry. It forms double-helical thin filaments composed of two strands. Within these, actin assembles in a head-to-tail manner, similar to its bacterial homologues [Michie and Lowe (2006)]. Such an alignment basis has obvious analogies in a pole-to-pole alignment of magnetic structures. Additionally, ferrofluids under field influences have been shown to form helical structures to minimize repulsions [see  5.1.2 - 5.1.3 ]. 
\subsection{ Possible origin of gradients; decoupling from 'mound'; origins of multicellularity}
The cytoskeleton, at least in eukaryotes, is organized via transmitted internal or external spatial cues, reflecting the polar organization of the cell [Drubin (2000)]. Now, the directed diffusion of synthesized products within the hypothesized structures would presumably lead to their gradients formed within these confined hyperstructures. These gradients could have been coupled to nonequilibrium fluctuations via suitable far-from-equilibrium chemical reactions. In this regard, magnetically sensitive chemical reactions have been considered [Weaver et al (2000)] wherein orientation of reactants could influence reaction rates. Such couplings could have helped towards enabling the exit of the replicator from the 'mound', by allowing continuation of directed transport in the absence of a thermal gradient. For, beyond protective barriers (say by amphi-pathic molecules) for navigating through a hostile environment outside the 'mound', the Russell compartments (and their membrane-resident hypothetical clusters) needed to evolve ways and means for replacing the many pathways crucial for their function that were intimately linked to the energy-providing, gradient rich scenario. Energy rich molecules would presumably have played a critical role in the `mound decoupling' as they took over from the 'mound generator' the charge of pumping energy into the range of dissipative processes outlined in previous sections. Independence from the 'mound' would have required selection in favor of resident clusters, with a progressively decreasing functional dependence on iron sulphide. Keeping in view the multidimensional properties of these minerals, fulfilling a variety of roles, this would be an extremely complex task [see  5.3.2]. 
\par
Anyhow, the continued presence of magnetic elements in the liberated, organic membrane bound, species (as for example in structural roles) would also offer a magnetic basis for association of these primordial organisms. A proposal on these lines, citing the multicellular life displayed by magnetotactic bacteria as a pathway for the evolution of a multicellular prokaryote, was also made recently [Davila et al (2007)]. Still later, as mentioned above, the slow entry into environments poor in FeS minerals would presumably have created a pressure for other mechanisms to take over and further reduce such dependence on magnetic matter.

\subsection{  Life evolves}

Russell et al envisage life as having dawned by 3.8 Gyr based on sedimentary evidences for biological CO$_{2}$ fixation. They propose that a non-free living universal ancestor was initially confined to structured FeS precipitates. This diverged into replicating systems en route to free-living eubacterial and archaebacterial cells, arising from separate locations on a single submarine seepage site. The eubacteria came to discover "the gold mine on the ocean's surface" and learnt to harvest sunlight. The presence of stromatolites, indicating photosynthetic bioactivity, by 3.5 Gyr, could signify the early occurrence of these landmark events [Martin and Russell (2003)].
\par
Thus independent schemes for lipid biosynthesis could be invented on separate locations of the mound, with different initial conditions [see above, Martin and  Russell (2003)]. This would be happening in separately located hosts --- like differently situated `LUCA broths' (last universal common ancestor), awaiting the emergence of proto-archaea and proto-eubacteria. Later, as evolution in the new openings took on new twists, subsequent replicators emanating from the 'mound' and navigating to these niches, would have found themselves in comparatively changed environments, thus permitting more variations. Furthermore,  each type of dwelling would have its individual selection criteria for survival. (By far, this mirrors a 'copy-paste-vary/modify -select' mechanism, for variations are ubiquitous in biology : at the level of DNA, RNA, proteins, modules, networks etc.). 
\par
The transfer of regulatory powers to the genes would have presumably been slow but progressive. As Newman and Muller (2000) have argued,  in the pre-Mendelian era  there was more plasticity in the mapping of phenotype with the genotype; this took on more of a one-to-one basis when morphological plasticity was seen to show a decline, perhaps as yet another `robustness' enhancing strategy.

\section{   Potential of coherent hierarchially nested structure}

\begin{center}
\begin{minipage}{13cm}
{\small {\em
\begin{quote}
Amra shobai raja amader ei rajar rajottey, noiley moder rajar shoney milbey ki shottey \\
-Tagore
\end{quote}
}
(We are all kings in this kingdom of our king, else how would we get along with our king.)
}
\end{minipage}
\end{center}

Finally we briefly review the physical significance and consequences of nested quasiperiodic order; particularly its potential in allowing a coherent transition between the quantum and classical domains. Could it generate adaptive structures entrained with a capacity for quantum information processing even as it simultaneously handled an intricate web of interactions? We review possible decoherence evasion strategies briefly. The discussion concludes with the final section surveying the myriad roles played by magnetism today in biology. What implications might this have for a possible magnetic origin of life? 

\subsection{  Smaller length scales to macroscopic: coherent transition}

\subsubsection{ Hierarchial long range order}
At the microscopic level too, a quasicrystalline ordered basis offers several points of departure in the study of these hypothetical structures. Although studies hitherto have been carried out primarily on alloy systems, sparked off by the 1984 discovery of such phases by Schechtman et al, the possibility exists for the generic features of such an arrangement to have commonalities with similar patterns procured by other means. In fact, the connection between the diffraction spectra of quasicrystalline solids and the dynamical spectra of substitution tilings modelling them, was realized from the appeal of mathematical models based on self-similar substitution tilings such as the Penrose tiling, providing the necessary ingredients of aperiodicity and long range order [Frank (2007)]. And, the discovery of quasi-crystals with pentagonal symmetry once again rekindled interest in the outstanding phenomena of phyllotaxis by providing a illuminating link between the periodicities observed in animate and inanimate forms [Adler et al (1997)]. As pointed out by Hotton et al (2006) two simple geometrical rules, 1) equivalent or nearly equivalent units are added in succession and 2) the position of new units is determined by interactions with units already in place, underlie the symmetries of myriad natural patterns having the regularity of crystals. 

\subsubsection {Phyllotaxis}
Phyllotaxis, known even to Kepler, was first identified in plants : the numerical regularity with which rows of nearest neighbours in lattices of plant elements (leaves, scales, etc.) form two families of intersecting spirals, with right and left-handed threads called parastichies [Levitov (1991)]. Their numbers in the two families are often consecutive Fibonacci numbers (1, 1, 2, 3, 5, 8, 13, 21, $\ldots$). The divergence angle between the base of two elements, e.g. consecutive primordia, is generally close to the golden angle $\alpha   = 360 (2- \tau$) degrees, $\tau$  being the golden mean $(-1 + \sqrt{5})/2 \sim 1.608$. A self-organizing system capturing some of these features, was set up by Douady and Couder (1996). Drops of ferrofluid, representing primordia, were deposited with a tunable periodicity in the center of a dish in the presence of a vertical magnetic field with a weak radial gradient, maximal at the periphery. The radial motion of the drops effectively mimicked phyllotactic growth owing to the optimizing interplay between inter-drop repulsive forces and their periodic introduction, demonstrating the success of a dynamical system in applying the underlying physics as iterative principles. In fact, the repulsive interactions between sequentially arising/deposited elements in an energy minimized manner, and manifesting in the mathematically precise arrangement patterns of phyllotaxis are seen as being generic to countless phenomena from different disciplines [Levitov (1991); Adler et al (1997); Nisoli et al (2007) and references therein] : repulsive magnetic dipoles (mentioned above), galactic structures, innumerable biological structures from the molecular level such as polypeptide chains, DNA to macroscopic levels evident in myriad forms such as marine-life, proportions in morphological and branching patterns [see Dunlap (1997)], Bernard convection cells, stress-driven self-assembly, bunched crystalline ion beams, atmospheric flows, and flux lattices in layered superconductors. The ubiquitousness of this phenomena is compelling as it can be seen to occur repeatedly, from the subnano to cosmological scales. 

\subsubsection  {The underlying manifold of interactions}
Highlighting the role of the underlying manifold in interactions between particles, Nisoli et al (2007), discuss possible novelties of interactions if instead of a conventional flat space, a geometrically constrained one is chosen, such as interacting repulsive elements (e.g. magnetic dipoles) on a cylinder [Nisoli et al (2004)]. (In this context, one could also note the universal importance of repulsive interactions between hubs in fractal networks [Song et al (2006)]).
\par
 	 Magnetism is known to host a rich spectrum of phenomena [Lange (2002)], arising out of alignment (or anti-alignment) of spins. Thus, in another recent study, a spiral arrangement of magnetic moments in quasi-periodic two-dimensional lattice has also been reported due to an interplay of competiting forces, namely, antiferromagnetic superexchange coupling and ferromagnetic direct exchange coupling [Hwang et al (2000)]. In the present context, repulsive interactions (such as negative charge on nucleates) could lead to (phyllotactic -like) framboid patterns  (see 3.4). 

\subsubsection{  Iterative tools}
Selvam has argued for self-similar structures with quasicrystalline order as iterative principles for preserving coherence [Selvam (1998)]. Iterative methods are the main tools for handling non-linear dynamics of perturbations providing mechanisms for evolving scale-free structures where patterns in the nested architecture are duplicated on all length scales. In this way,  the microscopic quantum world gets connected to the macroscopic realm with self-similar structures arising out of deterministic chaos. This existence of long-range spatiotemporal correlations where a non-dimensional scale factor can relate short- and long-term fluctuation amplitudes is the hallmark of self-similarity, manifest as self-organized criticality in natural dynamical systems [Bak and Chen, 1991]. In a similar vein, we point out that the potential information processing power of such hypothesized structures could well lie in their nested arrangement, permitting iterative computations. 

\subsection{  Fractal lattice: link to quantum searches in coherent structures}

\subsubsection{ Biological search problems }
	The statement 'Necessity is the mother of invention' quite appropriately describes biological systems, for they are intrinsically adaptive. For example  a crisis, following the lack of a nutrient, could push the system into a search, as otherwise a halted network of reactions could bring it to its end. A new pathway, represented by a node, could come to its rescue. In turn, this route would continue to be acceptable by feedback, i.e., 
as long as it could successfully permit the flow of networked reactions and maintain the system far away from equilibrium. In fact, biological evolution itself can be taken as an outstanding illustration of a biological search problem, with divergences symbolized by tree nodes. Other prominent examples include adaptive mutations, the clonal Darwinian-like phase in the adaptive immune system, brain connections and protein folding. Indeed, the latter gives a direct link to the heat-shock proteins that occur ubiquitously in all domains of life. The chaperone role of these early evolved molecules, implemented via an apparent  confining effect of the cavity walls on protein-folding space, seems intriguing [Bukau  and Horwich (1998)].   

\subsubsection{  Quantum search by 'active front'}

Let us define a  'quantum decision front' as comprising  a set of elements interacting in a local or non-local manner. This could involve, say a molecule or its parts, and/or concerned parts of its neighbours. This active front of the system 'decides' as a quantum whole [c.f., McFadden and Al-Khalili (1999)]. In a classical search for the optimal path, say for connecting amino acids in protein folding, the global decision for the system would need to take into account conflicting forces acting on each component. Each retains its identity as it interacts with its respective environment and the system is obliged to follow (like a voting pattern), an optimal path agreeable to its individual members ; however,   the ensuing  dynamics  may lead to  frustrations  in local minima. Also background noise due to other non-relevant `environment' interactions would compound the decision-making  process. In a quantum search on the other hand, the set of elements on the active front would be dissuaded from interacting with the `environment',  and would instead `step in phase' with one another to form a superposition of states. Thus, the joint or `holistic' decision taken by the single unit or the set of coherent active front elements, naturally makes this relatively a  more efficient approach for conducting a search. In this manner,  quantum coherence in a system holds the key to accessing the wave-property of  its constituent nanoscale elements. Now  the two main questions are : 1) Does Nature employ quantum coherent processes ?  2) If so, how does it maintain coherence in its highly interactive mileu ? 
\par
	Today, experimental indications of such quantum signatures in biology are slowly trickling in [Engel et al (2007); Bieberich (2000) ; reviewed in Hagan et al (2002)], happily repudiating  the `not proven' status of quantum biology [Davies (2004a)]. What's more, new tools from AMO physics, namely,   femtosecond laser-based 2D spectroscopy and coherent control approaches (using shaped laser pulses  tailor-based on adaptive feedback for control-driven reactions),  have shown that phase relationships in bio-nanostructures, undoubtedly play a critical role in modulating the course of bio-reactions [Nagya et al (2006)]. 

\subsubsection{  Biology and quantum search algorithms }

It seems Nature has quietly been using these strategies all along. The perfect matching of the solutions of Grover's search problem of a marked item in an unsorted database, to the base-pairing logic of nucleic acids in transcription and translational themes was pointed out by Patel as  evidence of the nucleic acid code for optimizing  a quantum search algorithm [Patel (2001)]. A directed walk through a superposition of all possibilities to the final state underlies the 2-fold increase in  sampling efficacy of the quantum algorithm over its classical counterpart which at best permits a random walk. Similarly,  a quantum search for a marked item in an unstructured database of space dimension $d$, gives a square root speed-up over the corresponding classical algorithm, a feature that also holds for the respective nested versions [Cerf et al (1998)]. Unmistakably, nested searches for structured problems are indicated for fractal processes, such as protein folding  proceeding via different structural length scales, that are suggestive of quantum scaling laws [Davies (2004a)].

\subsubsection {Decoherence and its evasion }

Due to the drastically reduced computational cost of quantum algorithms [Farhi \& Gutman (1996); Roland and Cerf (2002)], and in view of Nature's inclination towards economy, it is tempting to invoke quantum phenomena for understanding myriad search/learning problems in biological systems. But the biggest stumbling block againnst this lies in the general perception of the entire biological mileu as a wet and warm, decoherence inviting, environment. This is because the relative phase information holds the key for allowing correlations between the system components, crucial for quantum information processing. To that end, the 'screening effect',  or 'cocooning'  structural mechanisms providing insulation against interactions with the environment,  have been proposed [Patel (2001) ; see also Davies (2004a, 2004b)]. 
Moreover,  co-existing decoherence evasion mechanisms presumably permit  active entry points  to quantum effects [Hagan et al (2002); Davies (2004a)]. According to Matsuno, the gradual release of energy stored in ATP by actomyosin ATPase, in a sequence of quanta $E_m $ over time intervals of $\Delta t_{m}$ , underlies the huge order of magnitude discrepancy between the observed time interval of hydrolysis of 1 molecule of $ATP \sim 10^{-2}$ sec ,  and that calculated by considering energy release of $E = 5\times 10^{-3}$ erg (7kcal/mol) from a singly emitted quantum, or $\hbar / E \approx  2 \times 10^{-15}$. The obtained values of $E_m \sim 2.2 \times 10^{-19}$ erg and $\Delta t_{m} \approx  4.5 \times 10^{-9}$ sec indicates therefore $2.2 \times 10 ^{6}$ number of coherent energy quanta release during one cycle of energy release from a single ATP molecule. In Kelvin scale,  each energy quantum $E_m$ amounts to $1.6 \times 10^ {-3}$ K associated with the actomyosin complex. This very low local effective temperature thus enables these molecular machines to act as heat engines for exploiting the thermodynamic gradient,  taking average physiological temperature as 310K. In addition, it offers a finite possibility of hosting quantum events, provided adequate infrastructure for protective `screening effects' are in place. 

\subsubsection{  Coherence inducing magnetism : relevance to hypothesized clusters }

The Matsuno group also reported  that the coherent alignment of induced magnetic dipoles in ATP-activated actomyosin complexes was  maintained over the entire filament,  even in the presence of thermal agitations causing rapid decoherence. The energy of the dipole-dipole interactions per monomeric unit of $1.1\times 10^{-22}$ Joule was far below the thermal energy per degree of freedom at room temperature [Hatori et al (2001)].
By analogy, here we point out that the presence of a magnetic field would help  exert a coherent influence and overcome decoherence inducing thermal effects, critical for enabling quantum mechanical events in the nano-scale greigite clusters. Plus, the primary problem of decoherence in  arguments [Davies (2004a)] for and against quantum processes in biology, seems to have a succour in the poor thermal conductivity of the quasi-periodic lattice. 

\subsubsection{  Co-existence of classical and quantum domains  }

The long range spatial coherence across the quasiperiodic lattice of the hypothetical magnetic clusters would presumably permit their residence on the brink of quantum and classical realms. In biology, the latter appear to play complementary roles, almost analogous to the 'mind' or `thinking'-like undisturbed search and 'body' or 'doing'-like networked interactions, respectively. The co-existence of the two domains is essential for invoking quantum effects in biology. Thus the need for making a quantum decision could arise in the face of crises (say lack of a nutrient; see subsubsection 5.2.1), almost as if halted networked interactions in say one subsystem, prompt the formation of a `quantum decision front'  with the rest of the system  plus environment, waiting as a `composite environment' to interact with a possibly  meaningful  path (make a 'measurement' of the quantum front). A fruitful interaction of  the 'composite environment' with one chosen path would mean a simultaneous collapse of the quantum superposition of alternative paths [McFadden and Al-Khalili (1999)]. The newly created path would join up for propagating with other similarly born networked interactions, for taking over and maintaining the system far away from equilibrium in the non-linear regime. Thus the interacting phase continues till the next challenge forces the system to seek help from the co-existing quantum domain. 
\par
While the magnetic field due to `mound' constituents could have had an overall stabilizing influence against decoherence events, this effect would presumably continue even after 'take off' from the mound environment, thanks to the magnetic field arising from elements inside the departed system (see  4.8). Irrespective of  the schemes for reducing dependence on the latter,  the evolution of these quasiperiodic structures would  have had access to  both classical and quantum domains. This coexistence offers a credible framework for co-evolution of both these  sectors, right at the very origins of life, thus adding weight to the contemplated possible signatures of quantum information processing in evolved biological systems of today [Khrennikov (2007)]. If life truly evolved from such a coherent organization, it could help  understand why living systems respond coherently to fields  such as exerted by chemical gradients, which are increasingly being viewed as critical players in cellular events [Harold (2005)]. 

\subsection{  'The Importance of Being Magnetic'}

\subsubsection{  'The Giant'}
Magnetism in its myriad manifestations, at different scales (quantum to cosmological), forms the basis of an incessantly exploding field of intense research with its technological/applicational ramifications adding to the immense complexity of this lively area [Skomski (2007); Givord (2005)]. Further, the similarities with other phenomena provide connections with other disciplines by sharing theoretical approaches, e.g. spin glasses (see below),  and applications such as  ferrofluid based neural network [Palm and Korenivski (2007)]. To say the least, the repeated appearance of fractal themes is compelling -- from magnetic critical phenomena to finer length scales where quasiparticle behaviour in a magnetic field can be explained by fractional quantum numbers [see Jain (2007); Goerbig et al (2004)]. The latter are  mathematically connected with i) Farey series elements, $F_n$, belonging to fractal sets,  and ii) the Hausdorf dimension $h$  which is identified as the $second$ fraction of the fractal sets in $F_n$ [da Cruz (2005)]. 

\subsubsection{  Connection to `Russell scenario' }
In postulating a magnetic origin of life, we make a connection of the 'Russell scenario' with this giant field : The co-existence of myriad facets which allow a parallel access to a vast array of possibilities, and at different scales ranging from classical to quantum. Apart from this, magnetism ropes in yet another veritable 'Hercules': the multi-faceted quasiperiodic lattice. Thus, for example, if the interplay between magnetic and other forces leads to self-assembly into a quasiperiodic lattice, then the geometry of the latter lays down rules for its electronic properties [Kroha et al (2003)], while also offering biologically useful physical properties, e.g.,  low frictional coefficient, allowing smooth interlayer diffusion processes,  or even information- rich aperiodic surfaces. These multi-dimensional possibilities at different scales, are an aspect of paramount importance keeping in view the known, (e.g. roles of positional information for DNA function),   and being contemplated, ( e.g. the role of a phase in the DNA helical structure [Kwon (2007)] ), multi-faceted properties of biomolecules. Thus, the cytoskeleton (microfilaments, intermediate filaments and microtubules) is being viewed as a tensed $ tensegrity $  fractal framework orchestrating events at different levels, from driving cellular mechanics to information processing [Ingber (2003a,b)]. The quasiperiodic arrangement in neuronal cytoskeletal microtubules, is also thought to provide  access to quantum coherent processes [Hameroff (1994)]. In fact, thermodynamic and kinetic studies even seem to suggest scale- invariant structures  allowing short-lived quantum coherent superposition of binding states [Bieberich (2000)]. 

\subsubsection{ Possible quantum domain ramifications }

The quantum mechanical `spin' holds the key to magnetic behaviour rendering magnetism with a unique potential for providing a connection between microscopic and macroscopic domains in a coherent manner. Also as discussed, the long range magnetic order in the fractal quasiperiodic lattice with poor thermal conductivity, offers the required ingredients for quantum mechanical processes. Further, the temperature of the ocean water meeting the hydrothermal exhalate, may have been much lower -- from 30$^\circ$C even down to 0$^\circ$C [Martin and Russell (2003)]. In the context of these hypothetical clusters one could perhaps consider the potential effects of magnetism at still finer length scales where the wave nature of particles or even quasiparticles, is manifest in their transport properties. Particularly, the manner in which a fractally modulated voltage,  generated as a consequence of the Fractional Quantum Hall Effect (FQHE), might influence conduction phenomena in the quasiperiodic lattice. In addition, there has  been some progress in the understanding of how a magnetic field can influence spectra of systems with extended electronic states [Chakrabarti (2005)]. Of course  more effort is needed for further insight in this direction.  

\subsubsection{Divergent cousins ? }

We make a brief review of  magnetism - related signatures (oriented assembly, frustration, quantum coherence, etc.) in biological systems. Thus, the magnetic anisotropy in proteins and polypeptides, attributed to the diamagnetic anisotropy of the planar peptide bonds, permits their oriented self-assembly in a magnetic field. In this context,  a magnetism assisted alignment basis for assembly of a number of fibrous biostructures has been demonstrated [Torbet and Ronziere (1984) and references therein]. Another long appreciated connection is the close mapping between the impact of frustration in magnetically disordered systems and bioprocesses, such as protein folding, which in fact allowed common approaches to be formulated for a number of related search problems [Fraenkel (1997); Hollander and Toninelli (2004 ); Stein (1996)]. Yet another association made recently between the cytoskeletal network and percolation systems [Traverso (2005)], recalls an important feature of magnetism,  viz.,  long range connectivity (such as offered by magnetic percolation clusters forming fractal networks [Itoh et al (2006)] ). 
\par
Then again,  magnetically responsive  quantum structures have been considered in some proposed mechanisms for providing an entry into quantum coherent processes (superposition, entanglement) in biology. In the 1970s, Frohlich proposed that the change in shape of a protein molecule could be enabled via an internal oscillating dipole placed strategically at its hinge point,  that  underwent a quantum transition. In addition, the cell membranes were envisaged to provide a basis for alignment of the dipoles, leading to a long range quantum coherent state [Frohlich (1968, 1975)]. The quantum coherence on small length scales between spins of molecules associated with ion channels, has been proposed by Bieberich (2001) as enabling quantum computations in the brain with a fractal activity profile. A quasicrystalline basis for microtubule structure is yet another proposed mechanism for enabling quantum coherence [Hameroff (1994)]. 
\par
If  the  above piecemeal suggestions converge towards a fuller picture, one may  envisage a natural link with the (hypothesized)  magnetically ordered coherent structures  as an 'ancestor' to these different manifestations of  magnetism related signatures. In this scenario, a replacement of its 'hardware' would have necessitated that the 'substitutes' be endowed with the very multi-dimensional properties (from the macro to the subatomic levels) it could offer by itself alone. This is in contrast to the idea of quantum processes as having arisen as a product of adaptive evolution [Doll and Finke (2003)]. In this context we also note that, other than the hitherto known 3d and 4f electron systems, the possibility of magnetic order arising out of unpaired p - electron - containing systems, has also been considered [Givord (2005)].

\subsubsection{  Influence across kingdoms }

	Finally the most compelling case for a magnetic origin of life, hypothesized here, comes from the ubiquitous presence of magnetic influences across the kingdoms. The primary use for magnetic structures  considered hitherto is for navigation sensing in bacteria, algae, protists, bees, ants, fishes, dolphins, turtles and birds [Kirschvink and Gould (1981); Winkelhofer (2005); (in a special case it was also found to give mechanical protection in a deep-sea snail [Posfai \& Dunin-Borkowski (2006)])]. In plants, despite a lack of systematic physiological analyses, evidence suggests that the weak geomagnetic field and strong magnetic fields affect the growth pattern, differentiation,  and in certain instances also the orientation of plants and fungi [Galland and Pazur (2005)]. Again,  ferromagnetic elements found in a variety of tissues in the human brain [Kirschvink et al (1992)],  have provided diagnostic implications [Brem et al (2006)]. Further, new sensitive detection methods using atomic spin magnetometers are increasingly facilitating the diagnostic detection of weak magnetic fields generated by organs such as brain and heart [Xia et al (2006)]. These developments should hopefully  pave the way for an understanding of how thermal noise effects are overcome [Kirschvink (1992)] in effecting such low-frequency field influences in biology. The biophysical mechanisms underlying magnetoreception,  and the nature of its possible receptors,  need  unravelling for insights into the magnetic influence on biological systems of today. Indeed,  if magnetism is a relic of the past,  it is yet  another facet of  FeS  that comprises  clusters of present-day enzymes,  linked to their rocky roots ( see subsection 1.1)  on the Hadean ocean floor.   

\section{  Conclusions and scope}

Russell et al have argued that life's cradle could have been already busy by 3.8 Gyr, and it evolved fast enough for a branch to have made it to the ocean surfaces by 3.5 Gyr, as implied by photosynthetic bio-signatures. Clearly, such a replicator needed extraordinarily fast search algorithms to have covered so much ground. This brings us to profound questions linked to the nature of life : \\
 (A) Is life only a classical self-organizing phenomenon? 
(B) Is it's complexity linked with the quantum domain, enabling efficient searches \textit{a la 'a quantum mind'} ?  \\
The search for such a connection in the gradient-  rich 'mound ' with magnetic constituents,   brings us to the possibility of magnetic- field - assisted     formation of magnetically ordered  structures  within the gel-phase of the 'Russell-membranes'. A feedback-basis is in- built in such ordering, while the nested structures can serve to implement iterative principles. Indeed, the potential of a quasiperiodic order in endowing the structures with biologically relevant properties (such as diffusion through surface-limited structures) provides the gradient- rich 'mound' a logical access to possibilities with striking similarities to biological phenomena: 'budding' of dynamically self-assembled structures, modes of template based copying, molecular motors, optical polarity, pre-RNA world, spatially directed, nested organization, all within a gel-like medium.   
\par
 The efficiency of the adaptive nature of biological systems, and the fractal organization of its elements , press for a coherent conection between the classical and the quantum domains that appear to have co-evolved over time,  to manifest in its present complexity. This paper provides a physical basis to link up the domains at the very origins of life - a link where a continuity is maintained between these realms via a scale-free quasiperiodic framework -- in terms of coherence (enhanced by the role of the magnetic field) of the nested hierarchy, all the way down to lower scales. For this it has freely drawn on the theoretical ideas and experimental findings of a host of scientists (listed in the bibliography). We recognize that this study  has not even scratched the surface of what presumably lies unfathomed,  but this should be reason enough to urge a formal exploration and interdisciplinary insights from not only biologists, geologists, engineers and chemists but also mathematicians, computer scientists and physicists.

\begin{ack}
This work was entirely financed,  with full  infrastructural support,  by Dr.Jean-Jacques Delmotte. Mr.Guy Delmotte gave computer assistance and Dr.Vineet Ghildyal helped with manuscript processing. GMD acknowledges the driving force of Dr.Anand K.Bachhawat in this project; it's course was transformed with Dr.Bani M.Sodermark's key suggestion  " look for a fractal lattice" !  
\end{ack}

\end{document}